\documentclass[12pt,preprint]{aastex}

\lefthead{HAN, Hwang, \& Ryu} 
\righthead{CORRELATIONS BETWEEN PLANETARY PARAMETERS}

\begin{document}
\title{Correlations Between Planetary Microlensing Parameters}

\author{Cheongho Han, Kyu-Ha Hwang, and Yoon-Hyun Ryu}
\affil{Department of Physics,
Chungbuk National University, Cheongju 361-763, \\
Republic of Korea; cheongho,kyuha,yhryu@astroph.chungbuk.ac.kr}


\begin{abstract}
Characterization of microlensing planets requires modeling of 
observed light curves including many parameters.  Studying the 
dependency of the pattern of light curves on the lensing parameters 
and the correlations between the parameters is important to 
understand how the uncertainties of the planetary parameters 
are propagated from other parameters.  In this paper, we show 
that despite the apparent complexity of the pattern of light 
curves of planetary lensing events, the correlations between 
the lensing parameters can be understood by studying how the 
parameters affect the characteristics of lensing light curves 
such as the height and width, the caustic-crossing time scale, 
and the location and duration of planetary perturbations.  
Based on analytic arguments about the dependency of light curve 
features on the parameters, we obtain the correlations for the 
two representative cases of planetary events.  We also demonstrate 
the applicability of the correlations to general planetary events 
by actually obtaining the correlations from modelings of light 
curves produced by simulations.
\end{abstract}



\section{Introduction}

A microlensing event occurs when an astronomical object (lens) is
closely aligned with the line of sight toward a background star
(source). Microlensing causes change of the source star brightness 
and the resulting light curve is characterized by its smooth 
variation \citep{paczynski86}. If the lensing object is a star 
and it contains a planet, the resulting light curve can exhibit 
a discontinuous signature of the planet on the smooth light curve 
of the primary-induced event, and thus microlensing can be used 
as a method to search for extrasolar planets \citep{mao91, gould92}.  
Microlensing is sensitive to planets that are generally inaccessible 
to other methods, in particular cool planets at or beyond the snow 
line, very low-mass planets, planets orbiting low-mass stars, 
free-floating planets, and even planets in external galaxies. 
Therefore, when combined with the results from other surveys, 
microlensing planet searches can yield an accurate and complete 
census of the frequency and properties of planets.  Since the 
first discovery in 2004, 9 microlensing planets have been reported
\citep{bond04, udalski05, beaulieu06, gould06, gaudi08, bennett08, 
dong09, sumi10, janczak10}.

Characterization of microlensing planets requires modeling of 
observed light curves.  This modeling process requires to include 
many parameters because the pattern of lensing light curves and
the signals of planets take different forms depending on the 
combination of these parameters.  Therefore, studying the dependency 
of the pattern of light curves on the lensing parameters and the 
correlations between the parameters help to understand how the 
uncertainties of the planetary parameters are propagated from 
other lensing parameters.  This also helps to establish observational 
strategies for better characterization of planets.  However, it appears 
that the correlations between the lensing parameters are very complex 
due to the enormous diversity of lensing light curves resulting from 
the combinations of the numerous parameters.

In this paper, we show that despite the apparent complexity of 
the pattern of light curves of planetary lensing events, the 
correlations between the lensing parameters can be understood 
based on the dependency of the characteristic features of lensing 
light curves on the parameters.  We provide the correlations for 
the two representative cases of planetary events.  We also 
demonstrate the applicability of the correlations to general 
planetary lensing events by actually obtaining the correlations 
from modelings of light curves produced by simulations.

\section{Parameters}

The microlensing signal of a planet is a brief perturbation to the 
smooth standard light curve of the primary-induced single-lensing 
event.  Therefore, the parameters needed to describe planetary 
lensing light curves are broadly divided into two categories.  The 
first set of parameters is needed to describe the light curve of a 
standard single-lens event produced by the star hosting the planet.  
These parameters include the closest lens-source separation normalized 
by the Einstein radius, $\beta$ (impact parameter), the time of the 
closest lens-source approach, $t_0$, the time required for the source 
to transit the Einstein radius of the lens, $t_{\rm E}$ (Einstein 
time scale), the flux from the lensed star, $F_{\rm S}$, and the 
blended flux, $F_{\rm B}$.  With these parameters, the single-lensing 
light curve 
is represented by 
\begin{equation}
F(t)=F_{\rm S} A(t)+F_{\rm B},
\label{eq1}
\end{equation}
where
\begin{equation}
A(u)={u^2+2 \over u(u^2+4)^{1/2}};\qquad
u(t)=\left[ \left( {t-t_0\over t_{\rm E}}\right)^2+\beta^2\right]^{1/2}.
\label{eq2}
\end{equation}
Here $u$ represents the lens-source separation normalized by the 
Einstein radius $\theta_{\rm E}$.  These parameters characterize 
the global shape of lensing light curves such as the height and width.

Besides the single-lensing parameters, additional parameters are 
needed to described the detailed structure of the perturbations 
induced by planets.  These parameters include the planet/star mass 
ratio, $q$, the projected star-planet separation normalized by the 
Einstein radius, $s$, and the angle between the source trajectory 
and the star-planet axis, $\alpha$.  Since planetary perturbations 
are produced in most cases by close approaches or crosses of source 
stars over caustics, an additional parameter of the source size 
normalized by the Einstein radius, $\rho_\star$, is needed to describe 
the deviation of the perturbation affected by the finite-source effect.  
From the combinations of the lensing parameters, the light curves of 
planetary events exhibit various patterns.

\section{Correlations}

\subsection{Single-Lensing Parameters}

The correlations between the single-lensing parameters can be found
by investigating how the global features of lensing light curves such 
as the height and width vary depending on the parameters.  The height 
of the light curve is determined by the combination of the impact 
parameter $\beta$ and the blended light ratio, $f_{\rm B}=F_{\rm B}/
F_{\rm S}$.  When affected by blended flux, the impact parameter 
estimated based on the apparent height of the light curve is related 
to the blended light ratio as
\begin{equation}
\beta_{\rm B}=\left[ 2(1-A_{\rm max,B}^{-2})^{-1/2}-2\right]^{1/2};\ \ 
A_{\rm max,B}={A_{\rm max}+f_{\rm B}\over 1+f_{\rm B}},
\label{eq3}
\end{equation}
where $A_{\rm max,B}$ is the apparent peak magnification.  Then, 
with the increase of $f_{\rm B}$, $A_{\rm max,B}$ decreases, and 
thus $\beta_{\rm B}$ increases, implying that as the blended light 
increases, the peak magnification appears to be lower, and the 
resulting impact parameter becomes larger.  Therefore, it is found 
that {\it the blended flux ratio $f_{\rm B}$ and the impact parameter 
$\beta$ are correlated}.

Blending affects not only the height but also the width of light 
curves.  The event time scale estimated from the width of the 
blended light curve, $t_{\rm E,B}$, differs from the true value, 
$t_{\rm E}$.  The relation between the two time scales is \citep{han99}
\begin{equation}
t_{\rm E,B}=t_{\rm E} \left( { \beta_{\rm th,B}^2-\beta^2 \over  
\beta_{\rm th}^2-\beta_{\rm B}^2} \right)^{1/2}.
\label{eq4}
\end{equation}
Here $\beta_{\rm th,B}$ and $\beta_{\rm th}$ represent the threshold 
impact parameters of the source trajectory for event detections with 
the presence and absence of blended flux, respectively.  Because of 
blending, the threshold magnification is increased by $A_{\rm th,B} 
=A_{\rm th}(1+f_{\rm B}) -f_{\rm B}$ and thus the corresponding 
threshold impact parameter is lowered by $\beta_{\rm th,B}=[2(1-
A_{\rm th,B}^{-2})^{-1/2} -2]^{1/2}$, implying that to be detected 
the source star of a blended event should approach the lens closer 
than an unblended event.  From equation (\ref{eq4}), it is found 
that the increase of blended flux causes the measured event time 
scale shorter.  Therefore, {\it the blended flux ratio $f_{\rm B}$ 
and the Einstein time scale $t_{\rm E}$ are anti-correlated}. Since 
the blended flux ratio is correlated with the impact parameter, 
while it is anti-correlated with the time scale, it is found that 
{\it the impact parameter $\beta$ and the time scale $t_{\rm E}$ 
are anti-correlated}.

\begin{figure}[th]
\epsscale{0.9}
\plotone{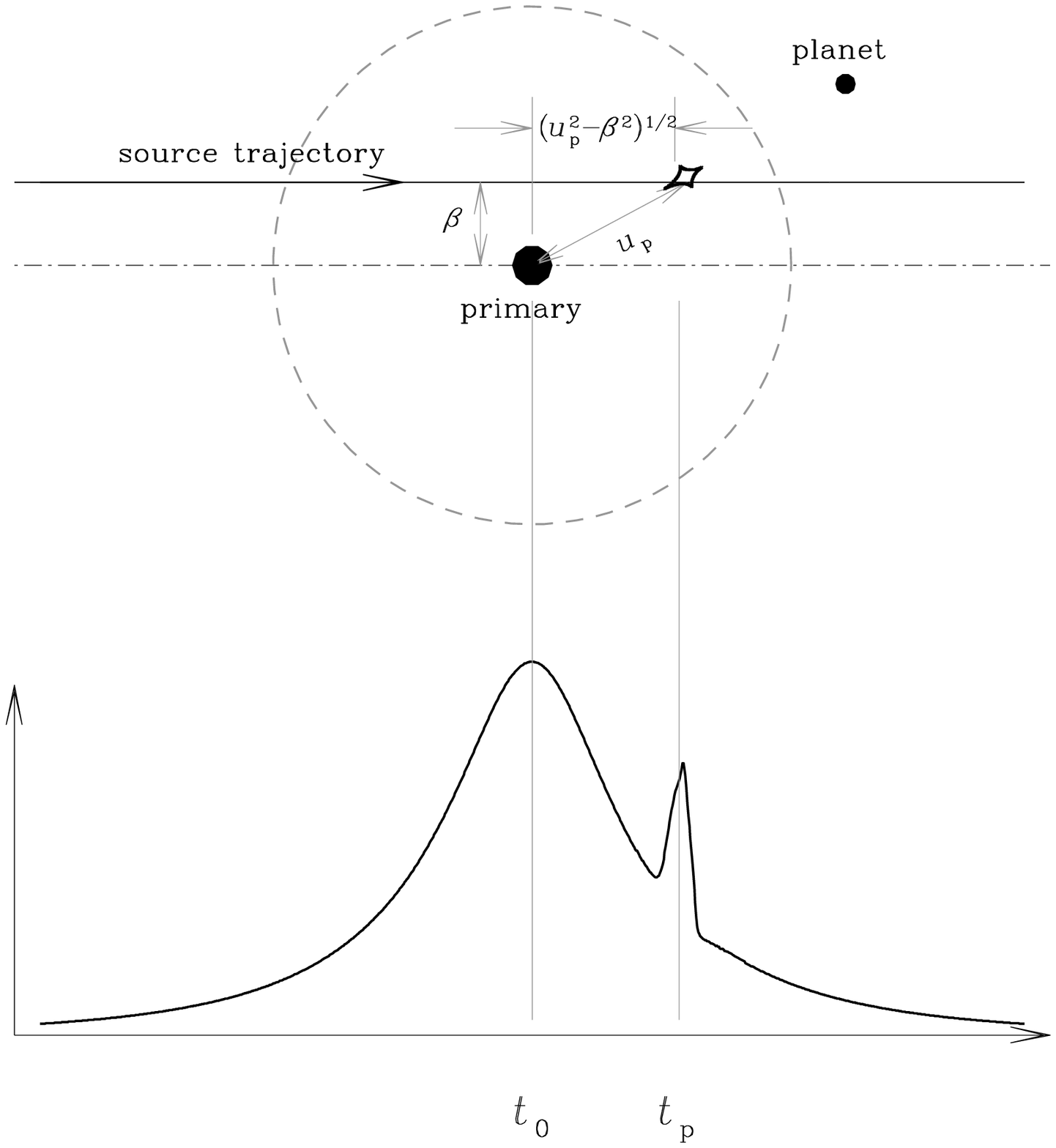}
\caption{\label{fig:one}
The geometry of the lens system showing the relation between the 
location of the planetary perturbation and lensing parameters.
The large dashed circle represents the Einstein ring and the 
small closed figure connected by concave curves is the caustic 
induced by the planet.
}\end{figure}

\begin{deluxetable}{cccccccccc}
\tablecaption{Correlations Between Lensing Parameters\label{table:one}}
\tablewidth{0pt}
\tablehead{
\multicolumn{1}{c}{} &
\multicolumn{1}{c}{$\beta$} &
\multicolumn{1}{c}{$t_{\rm E}$} &
\multicolumn{1}{c}{$\rho_\star$} &
\multicolumn{1}{c}{$s\ (s>1)$} &
\multicolumn{1}{c}{$q\ (s>1)$} &
\multicolumn{1}{c}{$s\ (s<1)$} &
\multicolumn{1}{c}{$q\ (s<1)$} 
}
\startdata
$\beta$      & $\nearrow$ &            &            &            &             &             &            \\
             &            &            &            &            &             &             &            \\
$t_{\rm E}$  & $\searrow$ & $\nearrow$ &            &            &             &             &            \\
             &            &            &            &            &             &             &            \\
$\rho_\star$ & $\nearrow$ & $\searrow$ & $\nearrow$ &            &             &             &            \\
             &            &            &            &            &             &             &            \\
$s\ (s>1)$   & $\nearrow$ & $\searrow$ & $\nearrow$ & $\nearrow$ &             &             &            \\
             &            &            &            &            &             &             &            \\
$q\ (s>1)$   & $\nearrow$ & $\searrow$ & $\nearrow$ & $\nearrow$ & $\nearrow$  &             &            \\
             &            &            &            &            &             &             &            \\
$s\ (s<1)$   & $\searrow$ & $\nearrow$ & $\searrow$ &            &             & $\nearrow$  &            \\
             &            &            &            &            &             &             &            \\
$q\ (s<1)$   & $\nearrow$ & $\searrow$ & $\nearrow$ &            &             & $\searrow$  & $\nearrow$ \\
             &            &            &            &            &             &             &                    
\enddata  
\end{deluxetable}

\subsection{Binary Parameters}

The major constraint on the the normalized planetary separation $s$ 
comes from the location of the planetary perturbation on the light 
curve.  This is because the location of the caustic with respect to 
the primary lens, ${\bf u}_{\rm p}$, is related to the position 
vector to the planet from the primary, ${\bf s}$, by \citep{griest98}
\begin{equation}
{\bf u}_{\rm p}= {\bf s}-{1\over {\bf s}}.
\label{eq5}
\end{equation}
The primary-caustic separation, $u_{\rm p}$, varies differently 
depending on whether the planet is inside ($s<1$) or outside ($s>1$) 
of the Einstein radius.  For planets with $s>1$, the caustic separation 
$u_{\rm p}$ increases with the increase of the planetary separation $s$.  
By contrast, the caustic separation decreases as the planet separation 
increases.  
In addition, the type of the perturbation differs depending on the 
location of the planet.  When the planet is located outside the 
Einstein ring ($s>1$), it perturbs the major image among the two 
images produced by the primary lens \citep{gaudi97}, resulting in 
higher magnifications than those of the adjacent part of the 
single-lensing light curve during the perturbation.  When the 
planet is inside the Einstein ring ($s<1$), by contrast, it perturbs 
the minor image and the magnifications during the perturbation are 
lower than those of the single-lensing light curve.
From the relation in equation (\ref{eq5}) combined with the geometry 
of the lens system presented in Figure \ref{fig:one}, it is found 
that the relative location of the perturbation on the light curve
is related to the time of perturbations, $t_{\rm p}$, 
and other lensing parameters by
\begin{equation}
{t_{\rm p}-t_0\over t_{\rm E}}=(u_{\rm p}^2-\beta^2)^{1/2}.
\label{eq6}
\end{equation}
Then, for planets with $s>1$, as $s$ increases, $u_{\rm p}$ increases,
and thus the impact parameter $\beta$ should increase to match 
the constraint given by the location of the perturbation on the 
observed light curve.  For planets with $s<1$, by contrast, 
$u_{\rm p}$ decreases with the increase of $s$, and thus $\beta$ 
decreases.  Therefore, it is found that {\it the planetary separation 
$s$ and the impact parameter $\beta$ are correlated for planetary 
events with major-image perturbations ($s>1$) and anti-correlated 
for events with minor-image perturbations ($s<1$)}.

The correlation between the planetary separation $s$ and the mass 
ratio $q$ is mainly given by the duration of the planetary perturbation.  
The duration is proportional to the size of the caustic.  The size of 
the caustic as measured by the width along the primary-planet axis 
is related to $s$ and $q$ \citep{han06} by
\begin{equation}
w_{\rm c}=
\cases{
4q^{1/2}/[s^2(1-s^{-2})]                    & for $s>1$, \cr
2q^{1/2} (\kappa-\kappa^{-1}+\kappa s^{-2}) & for $s<1$, 
\label{eq7}
}
\end{equation}
where $\kappa^2=[\cos 2\theta \pm (s^4-\sin^2 2\theta)^{1/2}]/
(s^2-s^{-2})$ and $\theta=\pi/2 \pm \sin^{-1}(\sqrt{3}s^2/2)/2$.
In the limiting cases of planetary separations of $s\gg 1$ and $s\ll 1$, 
the relations are expressed in compact forms of 
\begin{equation}
w_{\rm c}=
\cases{
4q^{1/2}s^{-2}                    & for $s\gg 1$, \cr
3\sqrt{3}q^{1/2}s^3/4             & for $s\ll 1$, 
\label{eq8}
}
\end{equation}
respectively.  Then, from the relation between $s$ and $q$ combined 
with the constraint of the perturbation duration measured from the 
light curve, it is found that {\it the planetary separation and the 
mass ratio should be correlated for planetary events with major-images 
perturbations ($s>1$) and anti-correlated for events with minor-image 
perturbations ($s<1$)}.

The main constraint on the normalized source radius $\rho_\star$ is 
provided by the duration of caustic-crossings (caustic-crossing time 
scale).  The caustic-crossing time scale is related to the normalized 
source radius $\rho_\star$ and the Einstein time scale $t_{\rm E}$ by
\begin{equation}
\Delta t_{\rm cc}={\rho_\star \over \sin\psi} t_{\rm E},
\label{eq9}
\end{equation}
where $\psi$ represents the angle between the caustic and the source 
trajectory.  Then, to match the constraint of the caustic-crossing 
time scale measured from the observed light curve, {\it the normalized 
source radius $\rho_\star$ and the Einstein time scale $t_{\rm E}$ 
should be anti-correlated}.

In Table \ref{table:one}, we summarize the correlations between the
lensing parameters.  Here, the correlations between the pairs of 
parameters not mentioned in the text are deduced from the correlations 
with other parameters.  For example, we deduce the correlation between 
the normalized source radius $\rho_\star$ and the impact parameter $
\beta$ based on the correlations between $\rho_\star$--$t_{\rm E}$ 
and $t_{\rm E}$--$\beta$ parameter pairs.  In the table, we mark 
``$\nearrow$'' for the pairs of parameters that are correlated, while 
the correlation is marked by ``$\searrow$'' for the pairs of parameters 
that are anti-correlated.

\citet{gould96} studied the relation between the normalized source 
radius and the impact parameter for high-magnification single-lens 
events.  See also \citet{gaudi97} for the relation between the 
lens-source proper motion, $\mu$, and the time scale, although 
$\mu$ is not a standard planetary lensing parameter.

\begin{figure}[th]
\epsscale{0.9}
\plotone{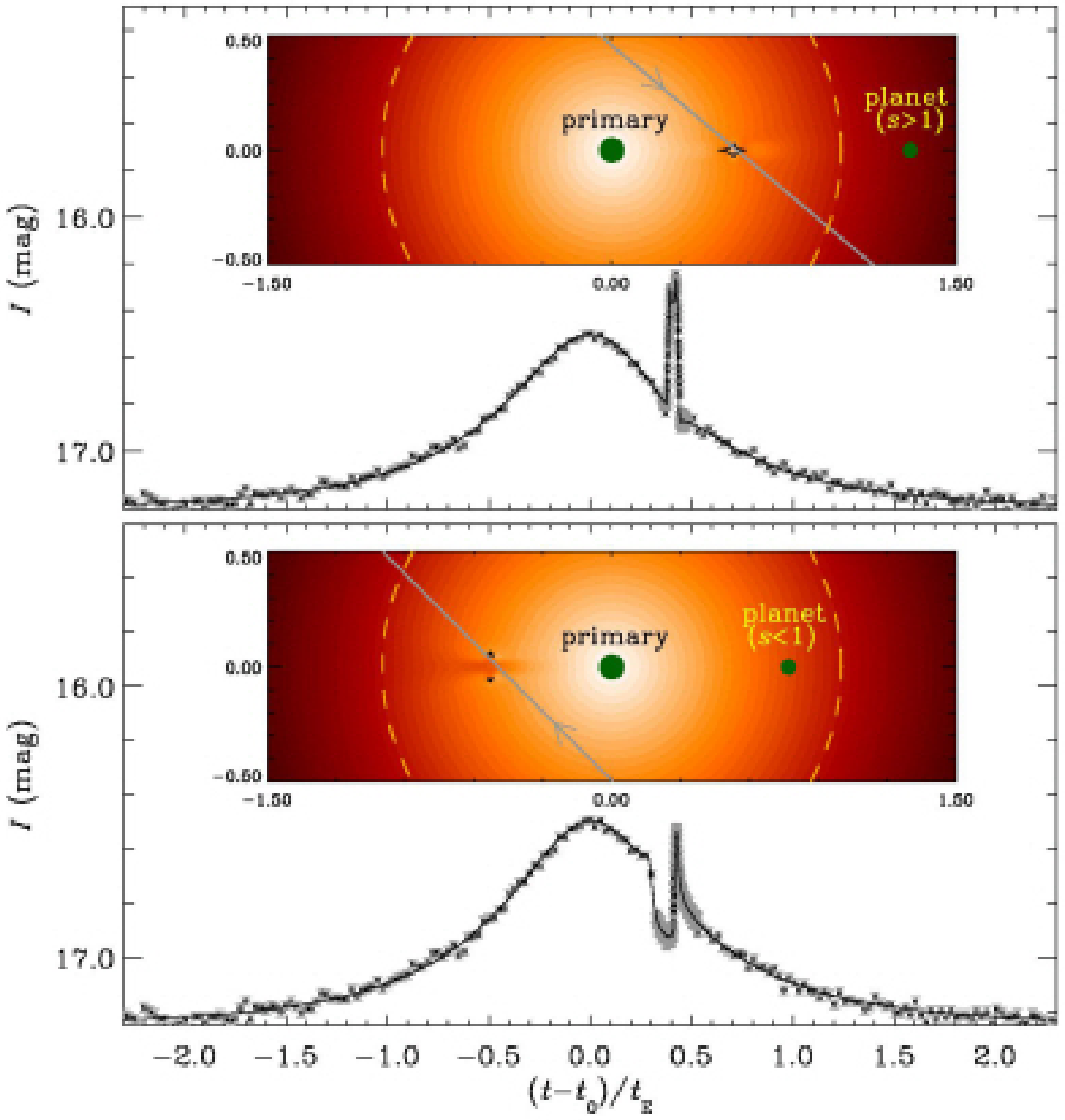}
\caption{\label{fig:two}
Light curves of planetary lensing events with major (upper panel) 
and minor-image (lower panel) perturbations.  The inset in each 
panel shows the geometry of the event where the locations of the 
primary and planet, source trajectory (straight line with an arrow), 
caustic (small closed figure), and Einstein ring (big dashed circle) 
are marked. The correlations between the lensing parameters of the 
individual events are presented in Fig.\ \ref{fig:three} and 
\ref{fig:four}, respectively.
}\end{figure}

\section{Demonstration}

In the previous section, we investigated the correlations between 
the lensing parameters based on analytic arguments about the 
dependency of the characteristic features of lensing light curves 
on the parameters.  In this section, we demonstrate that the 
correlations are applicable to general planetary microlensing 
events by actually obtaining the correlations from modelings of 
light curves produced by simulations.

We produce two light curves of planetary lensing events, where
the individual curves represent those of events with major and 
minor-image perturbations, respectively.  The light curves are 
produced considering the strategy of the current planetary lensing 
experiments where events are detected through modest-cadence 
survey observations and perturbations are densely covered by 
follow-up observations.  We set the photometric uncertainty by 
assuming that the photometry follows photon statistics with a 1\% 
systematic uncertainty and the deviations of the data points are 
Gaussian distributed.  The other factors affecting the photometry 
such as the source brightness and blending are based on the values 
of typical Galactic bulge events.  Figure~\ref{fig:two} shows the 
light curves of the events produced by the simulation where the 
upper and lower panels are those of the events with major and 
minor-image perturbations, respectively.  The inset in each panel 
shows the geometry of the event where the straight line with an 
arrow is the source trajectory and the temperature scale represents 
magnifications where a brighter tone implies a higher magnification.

\begin{figure}[t]
\epsscale{0.9}
\plotone{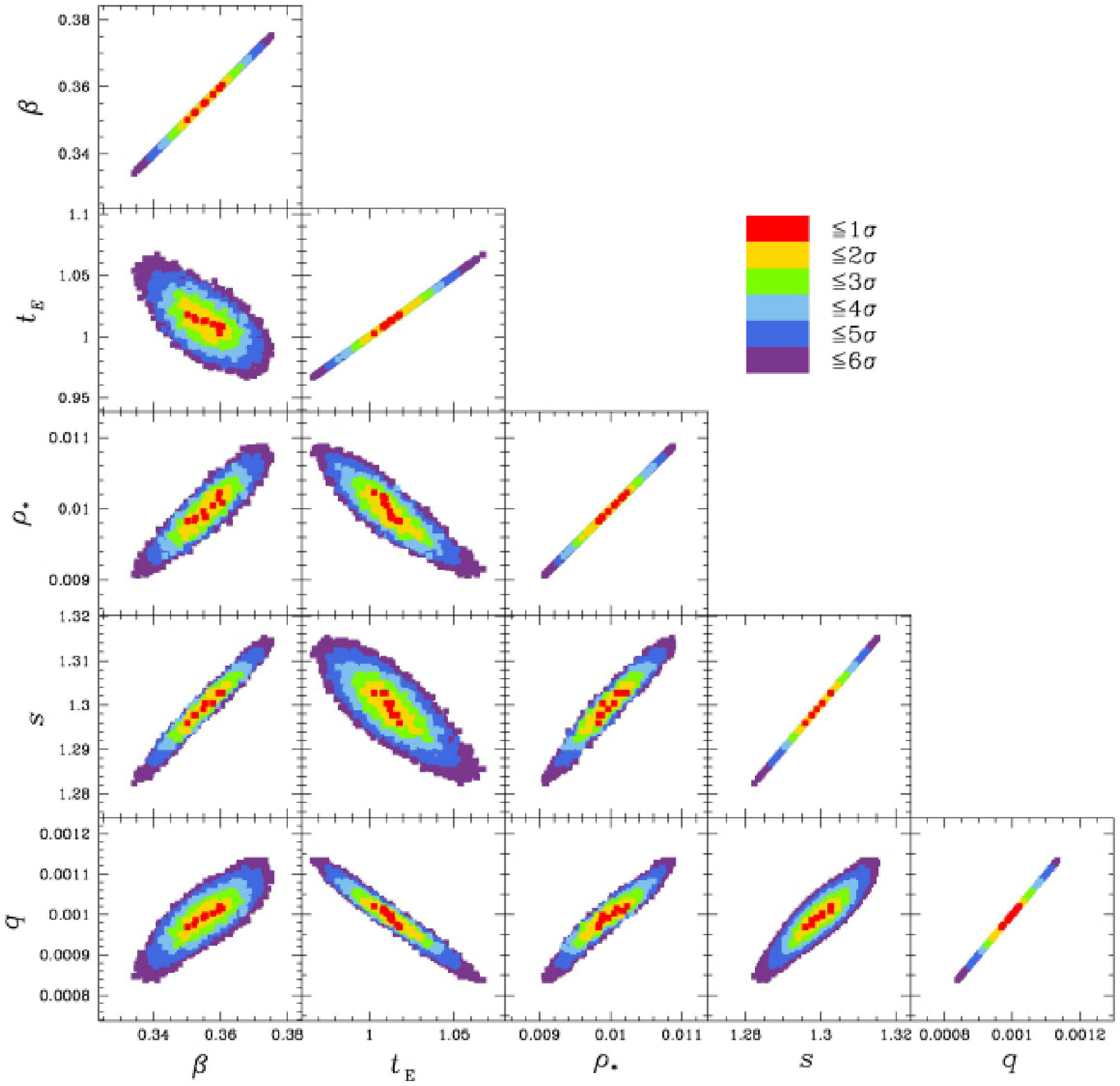}
\caption{\label{fig:three}
Contours of $\chi^2$ in the spaces of various combinations of
lensing parameters for the planetary event with a major-image
perturbations.  The result is based on the modeling of the light 
curve presented in the upper panel of Fig.\ \ref{fig:two}.
For direct comparison of the correlations to those presented 
in Table \ref{table:one}, the order of parameters are arranged 
accordingly.
}\end{figure}

\begin{figure}[t]
\epsscale{0.9}
\plotone{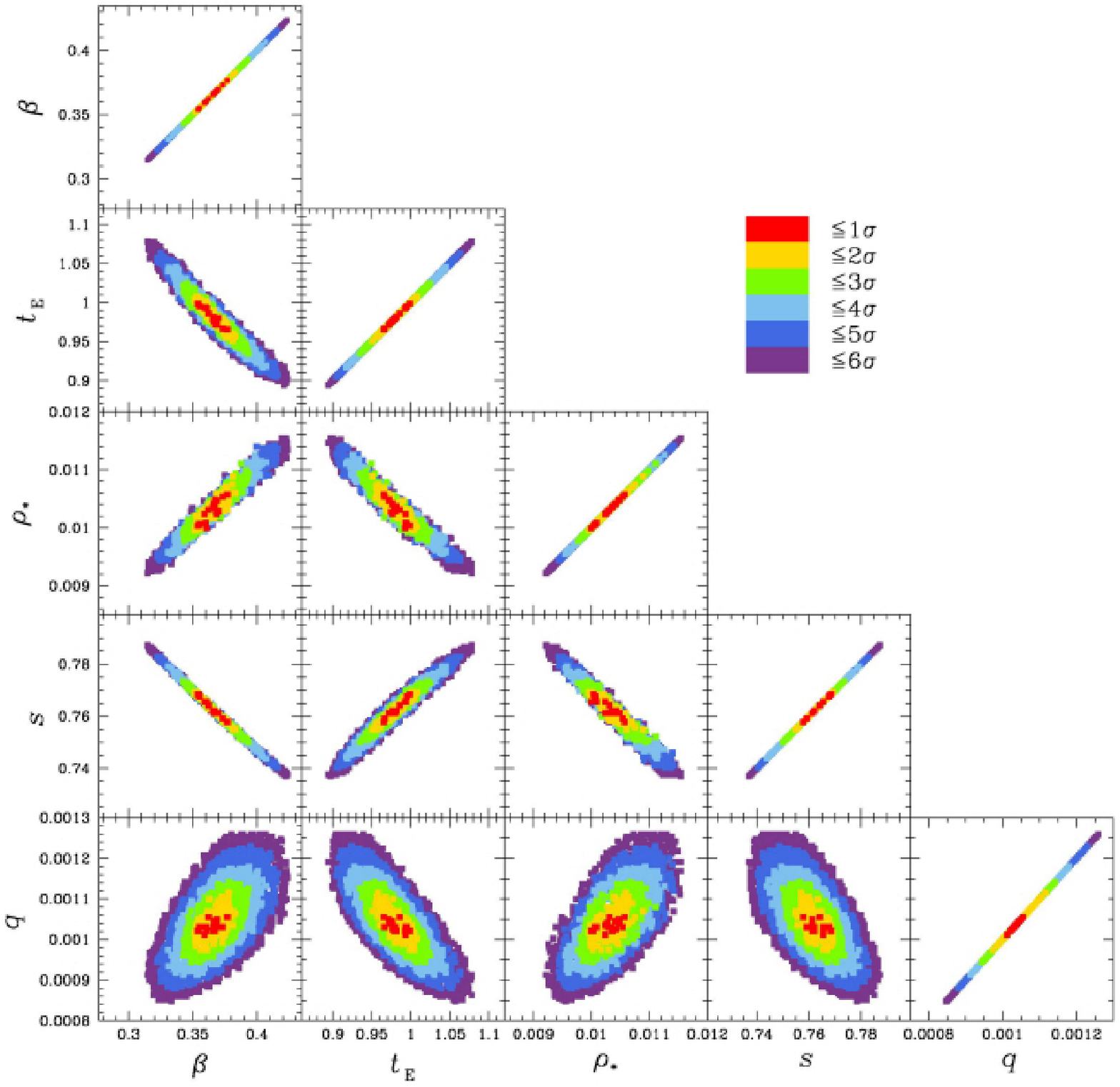}
\caption{\label{fig:four}
Same as in Fig.\ \ref{fig:three} except that the correlations 
are for the planetary event with a minor-image perturbation. 
}\end{figure}

We search for the solution of the lensing parameters by conducting 
modeling of the simulated light curves.  In modeling planetary 
lensing light curves, it is difficult to conduct brute-force search 
throughout all parameter space due to the large number of parameters.  
In addition, it is difficult to conduct simple downhill approach due 
to the complexity of the $\chi^2$ surface.  We, therefore, use a 
hybrid approach where grid searches are conducted over the space of 
parameters of $s$, $q$, and $\alpha$ and the remaining parameters 
are allowed to vary so that the model light curve results in minimum 
$\chi^2$ at each grid point.  We use a Markov Chain Monte Carlo method 
for $\chi^2$ minimization.  Once the solutions of the individual grid 
points are determined, the best-fit model is obtained by comparing 
the $\chi^2$ minima of the individual grid points.  The uncertainties 
of the best-fit parameters are estimated based on the chains of 
solutions produced by modeling.  For the computations of lensing 
magnifications including finite-source effect, we use the ray-shooting 
method where a large number of rays are shot from the image plane, 
bent according to the lens equation, and arrive on the source plane.  
Then, the magnification corresponding to the location of a source 
star is computed by comparing the number density of rays on the 
source star with the density on the image plane.  We minimize the 
computation time by restricting the region of ray shooting around 
the images and thus minimizing the number of rays needed for the 
computations of finite-source magnifications.  We also apply 
semi-analytic hexadecapole approximation \citep{pejcha09, gould08} 
for magnification computations in the region where the effect of 
source size is not important.

In Figure \ref{fig:three} and \ref{fig:four}, we present the results 
of the modeling for the planetary events with major and minor image 
perturbations, respectively.  The individual panels of each figure 
show the contour plots of $\chi^2$ in the spaces of the combinations 
of the lensing parameters.  To directly compare the correlations, 
we arrange the panels according to the same order of the parameters 
presented in Table \ref{table:one}.  From the comparison, it is 
found that the correlations found from modeling coincide with those 
predicted based on analytic arguments.  This implies that the 
correlations presented in Table \ref{table:one} are applicable to 
general planetary microlensing events.

\section{Conclusion and Discussion}

We investigated the correlations between the parameters of planetary 
lensing events.  From this, we found that  the correlations could be 
understood by studying how the lensing parameters affect the 
characteristics of lensing light curves such as the height and width 
of the light curve, the caustic-crossing time scale, and the location 
and duration of perturbations.  Based on analytic arguments about the 
dependency of the features of lensing light curves on the parameters, 
we obtain the correlations.  We also demonstrated the applicability 
of the correlations to general planetary lensing events by actually 
obtaining the correlations from modelings of light curves produced 
by simulations.

Understanding the correlations between lensing parameters can help 
to setup observational strategies for better constraints of the 
planetary parameters.  For example, the correlations between the 
blending and planetary parameters imply that blending affects 
determinations of the planetary parameters and thus de-blending or 
precise determination of the blending parameter is important to 
better constrain the planetary parameters.  Several methods can be 
applied to resolve the blending problem.  Photometrically, it is 
known that good coverage of the peak region and wings of light curves 
by follow-up observations help to constrain the blending parameter 
\citep{thomas06}. High resolution imaging from space observations 
\citep{han97} or ground-based AO imaging \citep{bennett10} can help 
to resolve source stars from blended stars. Astrometric measurement
of the blended image centroid can also help to identify the lensed
source among blended stars \citep{alard95}.
Precise and dense coverage of the perturbation is another way to 
constrain the planet parameters. This is because not 
only the location of perturbations but also their shape provides
constraints on planetary parameters.\footnote{The shape of 
perturbations is basically determined by the caustic shape, which 
depends on the planetary separation. Therefore, precise and dense 
coverage of perturbations is another way to improve the precision 
of planetary parameter measurements.  This is the same reason why 
it is possible to constrain the planetary parameters from the 
analysis of planetary perturbations occurring on high-magnification 
events. These perturbations occur near the peak of light curves 
regardless of the planetary separation. Nevertheless, it is possible 
to constrain the planetary parameters by analyzing the shape of 
perturbations.} Then, even if the location of the perturbation is 
uncertain due to severe blending, it is still possible to constrain 
the planetary parameters from the shape of the perturbation.

\acknowledgments 
This work is supported by Creative Research Initiative program
(2009-0081561) of National research Foundation of Korea.


\begin{thebibliography}{99}

\bibitem[Alard et al.(1995)]{alard95}
Alard, C., Mao, S., \& Guibert, J.\ 1995, \aap, 300, L17

\bibitem[Beaulieu et al.(2006)]{beaulieu06}
Beaulieu, J.\ P., et al.\ 2006, Nature, 439, 437

\bibitem[Bennett et al.(2008)]{bennett08}
Bennett, D.\ P., et al.\ 2008, \apj, 684, 663

\bibitem[Bennett et al.(2010)]{bennett10}
Bennett D.\ P., et al. 2010, \apj, 713, 837

\bibitem[Bond et al.(2004)]{bond04}
Bond, I.\ A., et al.\ 2004, \apj, 606, L155

\bibitem[Dong et al.(2009)]{dong09}
Dong, S., et al., 2009, \apj, 698, 1826

\bibitem[Gaudi \& Gould(1997)]{gaudi97}
Gaudi, B.\ S., \& Gould, A.\ 1997, \apj, 486, 85

\bibitem[Gaudi et al.(2008)]{gaudi08}
Gaudi, B.\ S., et al.\ 2008, Science, 319, 927

\bibitem[Gould(2008)]{gould08}
Gould, A.\ 2008, \apj, 681, 1593

\bibitem[Gould \& Loeb(1992)]{gould92}
Gould, A., \& Loeb, A.\ 1992, \apj, 396, 104

\bibitem[Gould \& Welch(1996)]{gould96}
Gould, A., \& Welch, D.\ L.\ 1996, \apj, 464, 212

\bibitem[Gould et al.(2006)]{gould06}
Gould, A., et al.\ 2006, \apj, 644, L37

\bibitem[Griest \& Safizadeh(1998)]{griest98}
Griest, K., \& Safizadeh, N.\ 1998, \apj, 500, 37

\bibitem[Han(1997)]{han97}
Han, C.\ 1997, \apj, 490, 51

\bibitem[Han(1999)]{han99}
Han, C.\ 1999, \mnras, 309, 373

\bibitem[Han(2006)]{han06}
Han, C.\ 2006, \apj, 638, 1080

\bibitem[Janczak et al.(2010)]{janczak10}
Janczak, J., et al. 2010, \apj, 711, 731

\bibitem[Mao, \& Paczynski(1991)]{mao91}
Mao, S., \& Paczynski, B.\ 1991, \apj, 374, L37

\bibitem[Paczy\'nski (1986)]{paczynski86}
Paczy\'nski, B.\ 1986, \apj, 304, 1

\bibitem[Pejcha \& Heyrovsk\'y(2009)]{pejcha09}
Pejcha, O., \& Heyrovsk\'y, B.\ 2009, \apj, 690, 1772

\bibitem[Sumi et al.(2010)]{sumi10}
Sumi, T., et al. 2010, \apj, 710, 1641

\bibitem[Thomas \& Griest(2006)]{thomas06}
Thomas, C.\ L., \& Griest, K.\ 2006, \apj, 640, 299

\bibitem[Udalski et al.(2005)]{udalski05}
Udalski, A., et al.\ 2005, \apj, 628, L109


















\end{thebibliography}
\end{document}